\documentclass[sigconf,screen]{acmart}

\AtBeginDocument{%
  \providecommand\BibTeX{{%
    Bib\TeX}}}

\copyrightyear{2024} 
\acmYear{2024} 
\setcopyright{rightsretained} 
\acmConference[ICPC '24]{32nd IEEE/ACM International Conference on Program Comprehension}{April 15--16, 2024}{Lisbon, Portugal}
\acmBooktitle{32nd IEEE/ACM International Conference on Program Comprehension (ICPC '24), April 15--16, 2024, Lisbon, Portugal}\acmDOI{10.1145/3643916.3644411}
\acmISBN{979-8-4007-0586-1/24/04}

\usepackage{amsmath,amsfonts}
\usepackage{graphicx}
\usepackage{textcomp}
\usepackage{xcolor}

\usepackage{algpseudocode}
\usepackage{algorithm}

\usepackage{etex}
\usepackage{booktabs}
\usepackage[utf8]{inputenc}
\usepackage[T1]{fontenc}
\usepackage{microtype}
\usepackage{graphicx}
\usepackage{paralist}
\usepackage{tabularx}
\usepackage{balance}
\usepackage{multicol}
\usepackage{multirow}
\usepackage{pbox}
\usepackage{enumitem}
\usepackage{lscape}
\usepackage{pifont}
\usepackage{xspace}
\usepackage{url}
\usepackage{tikz}
\usepackage{float}
\usepackage[TABBOTCAP]{subfigure}
\usepackage{ragged2e}
\usepackage{fontawesome}
\usepackage[figuresright]{rotating}
\usepackage{tcolorbox}

\usepackage{adjustbox}

\def\BibTeX{{\rm B\kern-.05em{\sc i\kern-.025em b}\kern-.08em
    T\kern-.1667em\lower.7ex\hbox{E}\kern-.125emX}}

  {\list{}{\leftmargin=0.1in\rightmargin=0.1in}\item[]}%
  {\endlist}

\newcommand{\ie}{\emph{i.e.,}\xspace}
\newcommand{\eg}{\emph{e.g.,}\xspace}

\newcommand{\etal}{\emph{et~al.}\xspace}

\newcommand{\figref}[1]{Fig.~\ref{#1}\xspace}

\newcommand{\tabref}[1]{Table~\ref{#1}\xspace}

\newboolean{showcomments}

\setboolean{showcomments}{true}

\ifthenelse{\boolean{showcomments}}
  {\newcommand{\nb}[2]{
    \fbox{\bfseries\sffamily\scriptsize#1}
    {\sf\small$\blacktriangleright$\textit{#2}$\blacktriangleleft$}
   }
  }
  {\newcommand{\nb}[2]{}
  }

\def\BibTeX{{\rm B\kern-.05em{\sc i\kern-.025em b}\kern-.08em
    T\kern-.1667em\lower.7ex\hbox{E}\kern-.125emX}}
\begin{document}

\title[On the Generalizability of Deep Learning-based Code Completion Across Programming Language Versions]{On the Generalizability of Deep Learning-based\\Code Completion Across Programming Language Versions}

%\author{Anonymous Authors\\
%Anonymous Institution\\
%}

\author{Matteo Ciniselli}
%\affiliation{STAR Group, Universit\`{a} della Svizzera italiana (USI), Switzerland}
\affiliation{
  \institution{SEART @ Software Institute, Universit\`{a} della Svizzera italiana (USI), Switzerland}
  \country{} 
}
\author{Alberto Martin-Lopez}
%\affiliation{SEART @ Software Institute, Universit\`{a} della Svizzera italiana (USI), Switzerland}
\affiliation{
  \institution{SEART @ Software Institute, Universit\`{a} della Svizzera italiana (USI), Switzerland}
  \country{} 
}
\author{Gabriele Bavota}
%\affiliation{SEART @ Software Institute, Universit\`{a} della Svizzera italiana (USI), Switzerland}
\affiliation{
  \institution{SEART @ Software Institute, Universit\`{a} della Svizzera italiana (USI), Switzerland}
  \country{} 
}

\renewcommand{\shortauthors}{Ciniselli et al.}

\begin{abstract}
Code completion is a key feature of Integrated Development Environments (IDEs), aimed at predicting the next tokens a developer is likely to write, helping them write code faster and with less effort.
Modern code completion approaches are often powered by deep learning (DL) models. 
% Deep learning (DL) models often power code completion plugins. However,
% It is common nowadays that code completion plugins are powered by deep learning (DL) models, trained on large amounts of code. However,
However, the swift evolution of programming languages poses a critical challenge to the performance of DL-based code completion models:
\emph{Can these models generalize across different language versions?}
This paper delves into such a question. In particular, we assess the capabilities of a state-of-the-art model, CodeT5, to generalize across nine different Java versions, ranging from Java 2 to Java 17, while being exclusively trained on Java 8 code. Our evaluation spans three completion scenarios, namely, predicting tokens, constructs (\eg the condition of an \texttt{if} statement) and entire code blocks.
The results of our study reveal a noticeable disparity among language versions, with the worst performance being obtained in Java 2 and 17---the most far apart versions compared to Java 8. We investigate possible causes for the performance degradation and show that the adoption of a limited version-specific fine-tuning can partially alleviate the problem. Our work raises awareness on the importance of continuous model refinement, and it can inform the design of alternatives to make code completion models more robust to language evolution.
\end{abstract}




\maketitle

% !TEX root = main.tex
%%%%%%%%%%%%%%%%%%%%%%%%%%%%%%%%%%%%%%%%
%%%%%%%%%%%%%%%%%%%%%%%%%%%%%%%%%%%%%%%%
\section{Introduction} \label{sec:intro}
%%%%%%%%%%%%%%%%%%%%%%%%%%%%%%%%%%%%%%%%
%%%%%%%%%%%%%%%%%%%%%%%%%%%%%%%%%%%%%%%%

% Introduction structure:
% - LLMs and DL-based approaches for software development.
% - Code completion task. One of the most popular and useful features. Started with symbolic or rule-based approaches, e.g., API suggestion, but nowadays mostly neural-based. Can be done at different levels, i.e., tokens, lines, blocks, etc.
% - Fast-paced evolution of programming languages: in the last 5 years, Java released new versions every 6 months, and Javascript and Python every year.
% - Problem statement and study setup: The impact of language evolution on the performance of DL-based code completion models has not been investigated yet. We investigate performance differences on code completion for Java.
% - Results, implications and future work: Differences with respect to older and newer versions. Small fine-tuning on the target version can improve performance. Raising awareness on the impact of retraining models. Towards online incremental training of models.

% - LLMs and DL-based approaches for software development.
Artificial Intelligence (AI) and, more specifically, Deep Learning (DL), have made their way into software engineering. Areas such as requirements engineering~\cite{amaral2023ml,dalpiaz2020requirements}, software design~\cite{ahmad2023towards,alomar2021we}, software development~\cite{nassif2019software,ernst2022ai} and software testing~\cite{mirabella2021deep,watson:icse2020} have seen the proposal of several DL-based solutions for their automation. 

In this context, Large Language Models (LLMs)~\cite{hou2023large} have recently gained popularity: LLMs are pre-trained on vast amounts of natural language and/or code and can be then fine-tuned to automate a specific task. LLMs are particularly well suited for tasks that can be formulated as a \emph{text-to-text transformation}~\cite{raffel:jmlr2019}, meaning that both the task input and output can be represented as text. For example, in \emph{code summarization}, the input text is a code snippet to document and the output text is its corresponding description in natural language. Many other tasks related to software development and testing can be formulated as text-to-text transformations, such as code review~\cite{tufano:icse2022}, code completion~\cite{ziegler:maps2022}, code refactoring~\cite{bhave2022deep}, test case generation~\cite{rao2023cat}, and bug-fixing~\cite{mastropaolo:tse2022}, among others.

% - Code completion task. One of the most popular and useful features. Started with symbolic or rule-based approaches, e.g., API suggestion, but nowadays mostly neural-based. Can be done at different levels, i.e., tokens, lines, blocks, etc.
Code completion is one of the most popular features in Integrated Development Environments (IDEs)~\cite{ciniselli2023source,svyatkovskiy:fse2020}. It aims at predicting the next tokens in a code snippet, thus it can save time and effort to developers. Research on code completion has evolved from traditional rule-based or symbolic techniques (\eg suggesting APIs to invoke according to the type of the variable that will store the returned value)~\cite{hou:rsse2010} to more sophisticated neural approaches which can make code predictions at multiple granularity levels going from a single code token to sections of code of an arbitrary size~\cite{wen:icse2021}. Thanks to their capabilities, DL-based code completion tools such as GitHub Copilot~\cite{copilot} (and its underlying Codex model~\cite{codex}) have pervaded the software industry, being nowadays used by millions of developers with a  positive effect on their productivity~\cite{peng2023impact}. 

% - Fast-paced evolution of programming languages: in the last 5 years, Java released new versions every 6 months, and Javascript and Python every year.
Just like code completion tools and approaches rapidly become more mature, programming languages themselves evolve at a fast pace too. In the last five years, languages such as Python~\cite{python} and JavaScript~\cite{javascript} have released major versions every year, and Java~\cite{java} has released a new version every six months. These new versions introduce new features and syntax such as new keywords, operators, data types, APIs and constructs. For instance, Java 8 introduced the \texttt{Stream} API, which allows developers to perform functional-style operations on collections of objects, and Java 11 introduced the \texttt{var} keyword, which allows developers to declare local variables without specifying their type. These new features may impact the performance of DL-based code completion models, which may be more prone to overfitting the specific language version used for their training, or may simply be unaware of new syntax constructs introduced in new language versions, thus being unable to predict them.
% Indeed, DL-based models are known to be particularly sensitive to the training data distribution~\cite{}, and the distribution of code snippets across different language versions may vary significantly. For example, the Java 8 version may be more represented in the training data as compared to the Java 17 version, thus the model may be more prone to predict code snippets written in Java 8 as compared to Java 17.

This problem is usually referred to as \emph{concept drift}~\cite{gama2014survey,lu2018learning}, where the data upon which a machine learning model has been trained on evolves over time, eventually leading to a performance degradation of the model or even invalidating it.

% - Problem statement and study setup: The impact of language evolution on the performance of DL-based code completion models has not been investigated yet. We investigate performance differences on code completion for Java.
In this paper, we study the impact of language evolution on the performance of DL-based code completion models. More specifically, we investigate to what extent code completion models can generalize across different language versions, including both older and newer versions as compared to the one used for training. 

We focus on Java as a good representative of a mature language with a long history of releases over the last 30 years.
The selected DL model for the evaluation is CodeT5~\cite{wang:emnlp2021}, a state-of-the-art code model based on the Text-To-Text Transfer Transformer (T5)~\cite{raffel:jmlr2019} architecture. This model has been shown to perform well in a variety of code-related tasks such as code summarization, code generation and defect detection, among others~\cite{wang:emnlp2021}. For our experiments, we pre-train and fine-tune CodeT5 exclusively on Java 8 code, and then assess its code completion capabilities on nine different Java versions (including Java 8 itself) ranging from Java 2 (released in 1998) to Java 17 (released in 2021). To make our study more comprehensive, we evaluate the model performance at three different code completion granularity levels, namely, token-level (\ie predicting the last $n$ tokens in a statement), construct-level (\ie predicting whole constructs such as \texttt{if} conditions), and block-level (\ie predicting all statements within a code block, such as the body of a \texttt{for} loop).

% - Results, implications and future work: Differences with respect to older and newer versions. Small fine-tuning on the target version can improve performance. Raising awareness on the impact of retraining models. Towards online incremental training of models.
The results of our study show significant performance differences across different language versions, with gradual decreases as we move away from the target version used for training---Java 8---and the worst performance being obtained in the most far apart versions---Java 2 and 17. This finding highlights the potential benefits of using version-specific DL models for code completion and, even more significantly, the importance of retraining the model on newer language versions as more training data is progressively available. As a matter of fact, we found that a small fine-tuning on the target language version can significantly improve the model performance, thus suggesting that the model can be adapted to new language versions with relatively little effort. 

Our work raises awareness on the importance of retraining DL models on new language versions, which can hopefully pave the way towards more effective code completion tools that developers can use to boost their productivity. Moreover, we believe that our results can be used to inform the design of online incremental training techniques for DL-based code completion models, which can be trained on new language versions as soon as they are released, thus keeping up with the fast-paced evolution of programming languages.

All code and data used in our study are publicly available \cite{replication}.
% !TEX root = main.tex
%%%%%%%%%%%%%%%%%%%%%%%%%%%%%%%%%%%%%%%%
%%%%%%%%%%%%%%%%%%%%%%%%%%%%%%%%%%%%%%%%
\section{Study Design} \label{sec:design}
%%%%%%%%%%%%%%%%%%%%%%%%%%%%%%%%%%%%%%%%
%%%%%%%%%%%%%%%%%%%%%%%%%%%%%%%%%%%%%%%%

\definecolor{gray50}{gray}{.5}
\definecolor{gray40}{gray}{.6}
\definecolor{gray30}{gray}{.7}
\definecolor{gray20}{gray}{.8}
\definecolor{gray10}{gray}{.9}
\definecolor{gray05}{gray}{.95}

\newlength\Linewidth
\def\findlength{\setlength\Linewidth\linewidth
	\addtolength\Linewidth{-4\fboxrule}
	\addtolength\Linewidth{-3\fboxsep}
}

\newenvironment{rqbox}{\par\begingroup
	\setlength{\fboxsep}{5pt}\findlength
	\setbox0=\vbox\bgroup\noindent
	\hsize=0.95\linewidth
	\begin{minipage}{0.95\linewidth}\normalsize}
	{\end{minipage}\egroup
	\textcolor{gray20}{\fboxsep1.5pt\fbox{\fboxsep5pt\colorbox{gray05}{\normalcolor\box0}}}
	\endgroup\par\noindent
	\normalcolor\ignorespacesafterend}

The \emph{goal} of this study is to assess the generalizability of a state-of-the-art DL-based code completion technique across different versions of the same programming language. In particular, we aim at answering the following research question (RQ):

\begin{center}	
	\begin{rqbox}
		\textit{To what extent do DL-based code completion techniques generalize across different language versions?}
	\end{rqbox}	 
\end{center}

Our empirical study is focused on the specific \emph{context} represented by: (i) CodeT5 \cite{wang:emnlp2021} as a representative DL model which has been used in the literature for the automation of code-related tasks \cite{chakraborty:fse22,bui:emnlp2022,troshin:emnlp22,zhou:icml23,le:neurips22,ahmed:tse2022,wang:acl22}; and (ii) code from 784 Java repositories for which we managed to reliably identify the used Java version. 

CodeT5 is a T5 model \cite{raffel:jmlr2019} pre-trained on code and natural language (\ie code comments). Wang \etal \cite{wang:emnlp2021} exploited the CodeSearchNet dataset \cite{husain:arxiv2019} for pre-training CodeT5. This dataset includes functions written in six programming languages (Go, Java, JavaScript, PHP, Python, and Ruby). Some of these functions also include a summary comment (\eg the Javadoc comment for Java). On top of CodeSearchNet, Wang \etal collected additional functions from C/C\# repositories hosted on GitHub. Overall, their pre-training dataset featured 8,347,634 functions, 3,158,313 of which paired with their documentation. The employed pre-training objectives were the classic masked language model (\ie self-supervised pre-training by randomly masking 15\% of the input asking the model to predict it) as well as a novel identifier-aware pre-training objective devised by the authors. 

In our study, we fine-tune CodeT5 for the code completion task \emph{only} with source code belonging to a specific Java version $v_i$, and we test it on multiple test sets, each featuring a different Java version (including $v_i$). The performance obtained on the $v_i$ test set provides a reference point for evaluating the performance on all the other test sets ($v_j \neq v_i$). To factor out the impact of pre-training on the CodeT5 model (whose pre-training dataset does include code from multiple Java versions~\cite{husain:arxiv2019}), we perform the experiments not only with the pre-trained model (\ie the publicly available CodeT5 checkpoint~\cite{codet5}), but also with a non pre-trained one.

\subsection{Data Collection and Datasets Creation}
\label{sub:dataset} 
In this section, we describe the process for collecting the data used to create the training and test sets. The training dataset is used to fine-tune CodeT5 for the code completion task on a single Java version, while the test sets are used to assess the generalizability of the model across different Java versions.

We used the tool by Dabic \etal \cite{dabic:msr2021} to select from GitHub all the non-forked Java repositories having more than 100 commits, 10 contributors, 50 issues, and 10 stars. We applied these filters in an attempt to remove toy/personal projects. This query resulted in 5,632 repositories. One of the requirements for our study is the ability to identify the Java version used in the project. For this reason, we only selected from the 5,632 repositories those using Maven. Indeed, Maven projects feature a \texttt{POM} (Project Object Model) file where developers can specify configuration details useful for building the project, including the used Java version. To increase the probability of collecting compilable code from these projects (\ie to avoid training the DL model on syntactically wrong code), we excluded all projects which cannot be compiled using Maven. In particular, we attempted the compilation for the last two GitHub releases of each repository, keeping only the ones for which compilation succeeded: we first check the latest release and, only if its compilation fails, we move to the second-last. 

Overall, we collected 784 repositories which can be successfully compiled and which explicitly report the adopted Java version in the \texttt{POM} file. From each repository, we randomly selected up to 1,000 Java files, ignoring the ones having the word ``test'' in the filename, aiming to exclude test files and create a more cohesive dataset made of production code only. The choice of capping the maximum number of files per project to 1,000 aims at avoiding that very large projects may contribute too much code to the final dataset. 

We used the set of collected files to build a method-level dataset of code completion tasks by slightly adapting the masking procedure proposed by Ciniselli \etal \cite{ciniselli:tse2021}. In particular, Ciniselli \etal proposed three method-level completion tasks having different levels of difficulty:

\begin{itemize}

\item \textbf{Token masking.} For each line of code, we masked the last $n$ tokens in a statement, with $n$ randomly ranging from three to ten, and then we ask the model to predict them. We chose to mask at least three tokens to avoid trivial completions (\eg only predicting the semicolon ending a Java statement). Moreover, we masked at most three random statements for each method to avoid generating too many instances from the same method. Indeed, each method can generate multiple training/testing instances, each with a specific statement being partially masked. The \emph{token masking} scenario emulates the code completion task in which the DL model is trying to complete the statement the developer is writing.\smallskip

\item \textbf{Construct masking.} Ciniselli \etal suggest that code completion performed on specific types of code constructs (\eg the condition of \texttt{if} statements) can be particularly useful to developers. We follow their \emph{construct masking} scenario, by masking all tokens used to implement: (i) the condition of an \texttt{if} statement or of a \texttt{while/for} loop, \eg ``\texttt{for(int i=0; i<dict.size(); i++)}'' is masked as ``\texttt{for(<MASK>)}'', with the model in charge of predicting the masked tokens; and (ii) the exception caught in a \texttt{catch} statement. Also in this case, a single method can contribute multiple instances to the dataset (\eg if it has three \texttt{if} statements, three instances are created, each of which having the condition of one \texttt{if} statement masked). We used srcML \cite{srcml} to identify the statements of interest and perform the masking.\smallskip

\item \textbf{Block masking.} This is the most challenging completion task. All code statements enclosed between curly brackets are considered a block (\eg the corpus of an \texttt{if} statement~\cite{ciniselli:tse2021}). As in the previous case, we used srcML \cite{srcml} to identify all blocks in each method and then create multiple instances of it, each having one entire block masked. As Ciniselli \etal, we only mask blocks featuring at most three statements to cap the task complexity.

\end{itemize}

We applied these masking procedures to all methods in the collected Java files, generating three different datasets. Exceptionally, we excluded methods that: (i) contained non-ASCII characters, which caused issues during training; and (ii) contained less than three or more than 50 statements (including signature), since they are either too short to prepare any meaningful completion scenario or too long to be provided as input to the DL model, whose maximum input length is 512 tokens.

We removed duplicate methods to avoid leaking of information between the training and test sets.
In the end, we collected 1,052,141 different methods, from which we derived a total of 2,846,746 \emph{token-masking} instances, 783,546 \emph{construct-masking} instances, and 1,303,444 \emph{block-masking} instances. Note that the \emph{construct-masking} instances are less than the overall number of methods since not all methods feature the construct types we mask (\ie \texttt{if}, \texttt{while}, \texttt{for}, and \texttt{catch}). 

The methods and corresponding instances are spread across 10 different Java versions as reported in \tabref{tab:dataset}.

% !TEX root = ../main.tex

% Design tables. 

\begin{table}[ht]
	\centering
	\small
	\caption{Number of methods and instances for each Java version.}
	\label{tab:dataset}
	\begin{tabular}{lrrrr}
	\toprule
	\bf Java & \multirow{2}{*}{\bf \# Methods} & {\bf \# Token} & {\bf \# Construct} & {\bf \# Block}\\ 
	\bf Version & & {\bf Instances} & {\bf Instances}& {\bf Instances} \\\midrule
 2 & 4,530 & 12,167 & 1,389 & 2,802\\
 5 & 8,549 & 21,178 & 6,510 & 11,103\\
 6 & 68,379 & 181,877 & 53,991 & 85,517\\
 7 & 99,405 & 266,162 & 72,108 & 122,593\\
 8 & 809,704 & 2,194,600 & 601,891 & 1,007,040\\
 9 & 4,809 & 13,879 & 3,393 & 5,401\\
 11 & 38,521 & 105,700 & 29,975 & 44,632\\
 14 & 6,160 & 17,203 & 5,842 & 8,644\\
 16 & 7,169 & 20,593 & 4,946 & 9,805\\
 17 & 4,915 & 13,387 & 3,501 & 5,907\\\midrule
 \bf ALL & \bf 1,052,141 & \bf 2,846,746 & \bf 783,546 & \bf 1,303,444\\
\bottomrule
\end{tabular}
\end{table}

\subsubsection{Creating the Test Sets}
\label{sub:test-sets}
Our goal is to create 30 different test sets, each one representing one of the 10 Java versions featured in our dataset and one of the three masking scenarios we adopted. For example, one test set will feature Java 2 methods in which entire blocks have been masked (\ie \emph{block masking}).
In creating the test sets, we must make sure that they all feature instances having a similar level of complexity, so that any observed performance differences in the model predictions can be attributed to the Java version used in the test set, and not to other factors.
To address this issue, we defined a number of metrics to assess the \emph{complexity} of each instance in our dataset, where an instance is a Java method having some of its tokens masked (based on the three masking scenarios previously described):

\begin{enumerate}
\item \emph{Number of lines in the method}. Longer methods may provide more contextual information to the model and allow for a simpler completion (\eg completing 10 out of 100 tokens may be easier than completing 10 out of 15 tokens).

\item \emph{Average number of characters per line}. Very long statements in a method may suggest a higher complexity of its instructions. The average number of characters per line is computed as the total number of characters in a method divided by the number of lines it features.

\item \emph{Number of masked characters}. The higher the number of masked characters, the higher the complexity of the code completion task (\ie guessing 20 characters is likely more complex than guessing five characters).

\item \emph{Number of lines masked (only for block masking)}. Similarly to the number of masked characters, masked blocks featuring a higher number of lines are likely more challenging to predict.
\end{enumerate}

% !TEX root = ../main.tex

% Design tables. 

\begin{table}[ht]
	\centering
	\caption{Complexity metrics computed on the dataset.}
	\vspace{-0.3cm}
	 \small
	% \scriptsize
	\label{tab:metrics}
	\begin{tabular}{lr}
	\toprule
	{\bf Metric} & {\bf Value} \\\midrule
Mean \# of lines in method & 7.7\\
%Median \# of lines in method & 4.0\\
Mean \# of characters per line & 28.1\\
%Median \# of characters per line & 26.0\\
Mean \# of masked characters (\emph{token masking}) & 18.2\\
Mean \# of masked characters (\emph{construct masking}) & 27.9\\
Mean \# of masked characters (\emph{block masking}) & 37.5\\
Mean \# of masked lines (\emph{block masking}) & 1.3\\
\bottomrule
\end{tabular} 
\end{table}
% !TEX root = ../main.tex

% Design tables. 

\begin{algorithm}
\scriptsize
\caption{Algorithm used for generating the test sets.}\label{algorithm}
	\begin{algorithmic}

\State $n \gets 5000$ \Comment{Number of selected methods for that specific Java version}
\State $\delta \gets 0.05$ \Comment{Difference allowed between the reference metrics and the test set metrics}

\Function{BuildTestSets}{$n,\delta$}

    \State $S \gets$ \Call{RandomlySelectSample}{$n$}
    
    \For{${steps} \gets 1$ to $5000$}
        \State $constraintMet \gets$ $True$
        \For{$i \gets 1$ to $NumberOfMetrics$}
            \State $R_{C_i} \gets$ \Call{ReferenceValueForMetric}{$i$}
            \State $S_{C_i} \gets$ \Call{CalculateSampleMetric}{$S, i$}
            
            \If{$S_{C_i} - R_{C_i} > \delta$}
                \State $constraintMet \gets$ $False$
                 %\State \textbf{Select} a random method $m_j$ \textbf{from} $S$ \textbf{where} $C_i > R_{C_i}$
                \State $m_j \gets$ \Call{FindMethod}{$S, i, R_{C_i}$} \Comment{ random method \textbf{having} $C_i > R_{C_i}$}
                 \State $m_k \gets$ \Call{FindMethod}{$S, i, R_{C_i}$} \Comment{ random method \textbf{having} $C_i < R_{C_i}$}

                \State $S \gets$ \Call{ReplaceMethod}{$S, m_j, m_k$}
                    \ElsIf{$S_{C_i} - R_{C_i} < \delta$}

                \State $constraintMet \gets$ $False$
                \State $m_j \gets$ \Call{FindMethod}{$S, i, R_{C_i}$} \Comment{ random method \textbf{having} $C_i < R_{C_i}$}
                 \State $m_k \gets$ \Call{FindMethod}{$S, i, R_{C_i}$} \Comment{ random method \textbf{having} $C_i > R_{C_i}$}                
                 \State $S \gets$ \Call{ReplaceMethod}{$S, m_j, m_k$}
            \EndIf
        \EndFor
        
        \If{$constraintMet =$ $True$}
            \State \textbf{return} $S$
        \EndIf
    \EndFor
    
    \State $n \gets n - 500$
    \State \textbf{return} \Call{BuildTestSets}{$n,\delta$}
\EndFunction
\end{algorithmic}
\end{algorithm}

We computed these metrics for all collected methods. \tabref{tab:metrics} reports their mean values on the whole dataset. These represent our reference values that we use to build the test sets for the different Java versions. In particular, we build the test sets by adopting the algorithm reported in Algorithm~\ref{algorithm}, explained in the following. 

We target the building of test sets featuring $n$=5,000 instances each (\eg 5k instances in the Java 8 test set using \emph{token masking}). We set as constraint that each test set should feature instances being close, in terms of complexity, to the reference metric values in \tabref{tab:metrics}, with a margin $\delta$ of $\pm$5\% for each metric. This means, for example, that a test set can have a mean number of lines per method being $7.7\pm0.38$. The same holds for the other metrics in \tabref{tab:metrics}. The algorithm starts by randomly selecting a sample $S$ of 5,000 Java methods from the dataset, checking for each metric $C_i$ how far this sample is from the reference value. If the constraint for $C_i$ is satisfied (\ie $S_{C_i}$ is within $\delta$ from the reference value $R_{C_i}$), no changes to $S$ are performed. Otherwise, if $S_{C_i}$ - $R_{C_i}$ $>$ $\delta$ and it is a positive number (\ie methods in $S$ have a higher value for the metric $C_i$ as compared to the reference value $R_{C_i}$), we remove from $S$ a randomly selected method $m_j$ having $C_i$ $>$ $R_{C_i}$ and we add a randomly selected method $m_k$ having $C_i$ $<$ $R_{C_i}$. 

If, instead, methods in $S$ have a lower value for the metric $C_i$ as compared to $R_{C_i}$, we remove from $S$ a randomly selected method $m_j$ having $C_i$ $<$ $R_{C_i}$ and we  add a randomly selected method $m_k$ having $C_i$ $>$ $R_{C_i}$. Essentially, we seek for convergence of the considered metrics. We perform this procedure for a maximum of 5,000 steps. If after 5,000 steps the sample $S$ does not meet the constraint for all metrics, we reduce the number of methods to collect by 500 (\ie $n = n - 500$) and repeat the process.

% !TEX root = ../main.tex

% Design tables. 

\begin{table}[ht]
	\centering
	\caption{Number of methods and instances for each set for each Java version.}
	 \scriptsize
	\label{tab:datasets}
	\begin{tabular}{llrrrr}
	\toprule
	\multirow{2}{*}{\bf Dataset} & \bf Java & \multirow{2}{*}{\bf \# Methods} & {\bf \# Token} & {\bf \# Construct} & {\bf \# Block}\\ 
	& \bf Version & & {\bf Instances} & {\bf Instances}& {\bf Instances} \\\midrule

train & 8 & 724,130 & 1,962,673 & 538,427 & 900,927\\\addlinespace[0.08cm]\hline\addlinespace[0.08cm]

eval & 8 & 80,574 & 218,377 & 59,729 & 99,886\\\addlinespace[0.08cm]\hline\addlinespace[0.08cm]
	
 \multirow{9}{*}{test} &  2  & 3,000 & 7,692 & 879 & 2,052\\ 
 & 5  & 5,000 & 12,626 & 4,670 & 6,866\\ 
 & 6  & 5,000 & 13,272 & 4,032 & 6,394\\ 
 & 7  & 5,000 & 13,440 & 3,547 & 6,069\\ 
 & 8  & 5,000 & 13,550 & 3,735 & 6,227\\ 
 & 11  & 5,000 & 13,547 & 3,494 & 5,650\\ 
 & 14  & 5,000 & 13,855 & 4,178 & 6,750\\ 
 & 16  & 5,000 & 14,356 & 3,568 & 6,849\\ 
 & 17  & 4,000 & 10,894 & 2,925 & 4,834\\
\bottomrule
\end{tabular}
\vspace{-0.2cm} 
\end{table}

Following this procedure, we managed to collect 5,000 methods for all test sets with few exceptions: for Java 2, which featured 4,530 overall methods in the whole dataset, we managed to collect 3,000 methods. For Java 17, we collected 4,000. Lastly, for Java 9, we did not reach convergence at any test set size, hence its exclusion from our study.

\subsubsection{Creating the Training (Fine-Tuning) and Evaluation Set}
\label{sub:training-evaluation-sets}
Given the distribution of methods across the Java versions in our dataset (see \tabref{tab:dataset}), we decided to use Java 8 as the ``training version'', since it features $\sim$80\% of the overall methods we mined. This means that Java 8 is the version on which we train CodeT5 to then test its performance on the test sets belonging to the different versions (including Java 8 itself). To create the training and evaluation sets, we took the 804,704 Java 8 methods which were not included in the Java 8 test set and split them into training (90\%) and evaluation (10\%) sets, the latter being used for hyperparameters tuning (as described in the next section). \tabref{tab:datasets} reports the number of methods and instances for the training, evaluation and test sets.

\subsection{Hyperparameters Tuning and Training}
We adopt the default parameters used in the paper presenting CodeT5 \cite{wang:emnlp2021}, only experimenting with different learning rates. More specifically, we evaluated three different values (\ie $1e^{-5}$, $2e^{-5}$, $5e^{-5}$), using the AdamW optimizer \cite{loshchilov:iclr19} to update the weights. We trained the model for 10K steps using a batch size of 12 and we evaluated the performance of each of the three configurations on the evaluation set in terms of percentage of exact match predictions (\ie the predicted code tokens are identical to the masked ones). We performed this tuning for both: (i) a CodeT5 model fine-tuned based on the publicly available pre-trained checkpoint \cite{codet5}; and (ii) a CodeT5 fine-tuned from scratch, not pre-trained at all. Indeed, as previously explained, CodeT5 has been pre-trained on source code, including Java code. 

Thus, it has seen Java code written in versions different from the Java 8 we want to use as ``training version''. Experimenting also with a non pre-trained model in our study allows us to factor out this further confounding factor. 

For both models (\ie pre-trained and non pre-trained) we found the best configuration to be the one in \tabref{tab:hyper}. Using this configuration, we trained the two models for 15 epochs, for a total of 4,250,000 training steps with a batch size of 12. The trainings took 26 days using an NVIDIA GPU GeForce RTX 3090 with 24GB of RAM. 

\begin{figure}[tb]
    \centering
    \includegraphics[width=\linewidth]{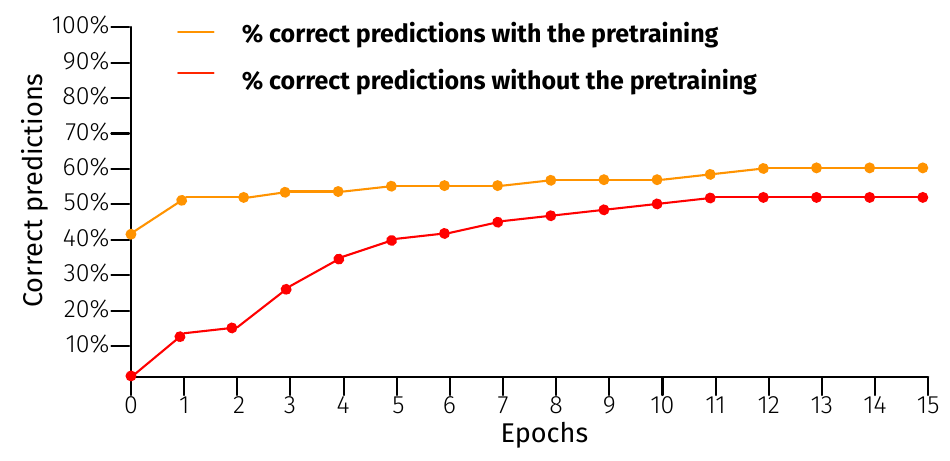}
    \caption{Percentage of Exact Match predictions for both models at different training epochs.}
    \label{fig:convergence}
\end{figure}

During the training, we saved a checkpoint every 50K steps, evaluating the performance on the evaluation set. We aimed to use early stopping, however, in both trainings, the best checkpoint was the last one for the pre-trained model and the second-last for the non pre-trained one. We are aware that this indicates potential margin for improvements for both models. However, the increment in performance among the latest checkpoints was very minor as it can be seen in \figref{fig:convergence}. Thus, we decided to not invest additional time in further training the models.

% !TEX root = ../main.tex

% Design tables. 

\begin{table}[ht]
	\centering
	\caption{Configuration used for the CodeT5 training.}
	 \small
	\label{tab:hyper}
	\begin{tabular}{lr}
	\toprule
	{\bf Hyperparameter} & {\bf Value}\\\midrule
	Learning rate & $5e^{-5}$\\
	Batch size &  12\\
	\# Encoder block & 12\\
	\# Decider block & 12\\
	\# Attention Heads & 12\\
	Hidden Layer Size  & 768\\
\bottomrule
\end{tabular} 
\end{table}

\subsection{Evaluation and Analysis}
\label{sub:datacollection}
We evaluated both trained models (\ie with and without pre-training) on all the 27 test sets (\ie 9 Java versions $\times$ 3 masking scenarios), collecting their predictions. For each test set, we compute the percentage of Exact Match (EM) predictions. We focus our discussion on the gap of performance (if any) existing between the EMs observed on the Java 8 test set (featuring instances written using the same language version used for the training and hyperparameters tuning) and those obtained on other Java versions, both those preceding and following version 8. 

We statistically compare the results achieved on the Java 8 test set and on all other test sets assuming a significance level of 95\%. We compute the Fisher's exact test (and related OR) on a matrix containing, for the different test sets and for different masking levels, the number of correct (EM) and incorrect predictions. To account for multiple test instances---we contrast the performance on Java 8 with all other versions---we adjust \emph{p}-values using the Benjamini-Hochberg procedure \cite{yoav:jstor1995}.
To bolster our findings, we complement our analysis with the CrystalBLEU score \cite{eghbali:icse2022} between the predictions made on the different test sets and the expected predictions. The CrystalBLEU score is computed in a similar way to the BLEU score~\cite{papineni:acl2002}, but the most frequent sequences of words are ignored, which are mostly due to the syntactic constructs and coding conventions of the programming languages~\cite{eghbali:icse2022}.
Specifically, we generated the boxplots of the distribution of the CrystalBLEU for each Java version, to evaluate whether comparable conclusions could be drawn with respect to those based on the EM predictions.
Although several metrics have been proposed to evaluate the quality of code and language model predictions (\eg BLEU \cite{papineni:acl2002}, ROUGE \cite{lin:tsbo2004} and CodeBLEU \cite{ren:arxiv2020}), we relied on EM predictions since they are widely used in the literature even for the prediction of entire blocks of code \cite{alon:icml2020,asaduzzaman:icsme2014,Ciniselli:msr2021,ciniselli:tse2021,hellendoorn:fse2017}, and CrystalBLEU since, according to its authors, it can distinguish similar from dissimilar code 1.9--4.5 times more effectively when compared to BLEU, since it is not inflated by the syntactic sugar of programming languages~\cite{eghbali:icse2022}.

% !TEX root = main.tex
%%%%%%%%%%%%%%%%%%%%%%%%%%%%%%%%%%%%%%%%
%%%%%%%%%%%%%%%%%%%%%%%%%%%%%%%%%%%%%%%%
\section{Results Discussion} \label{sec:results}
%%%%%%%%%%%%%%%%%%%%%%%%%%%%%%%%%%%%%%%%
%%%%%%%%%%%%%%%%%%%%%%%%%%%%%%%%%%%%%%%%
We discuss the achieved results by presenting: (i) performance differences across language versions; (ii) possible reasons behind the observed differences in performance; and (iii) the impact of version-specific fine-tuning on the model performance. The combination of these analyses allows to answer our RQ.

\begin{figure*}[tb]
    \centering
    \includegraphics[width=0.67\linewidth]{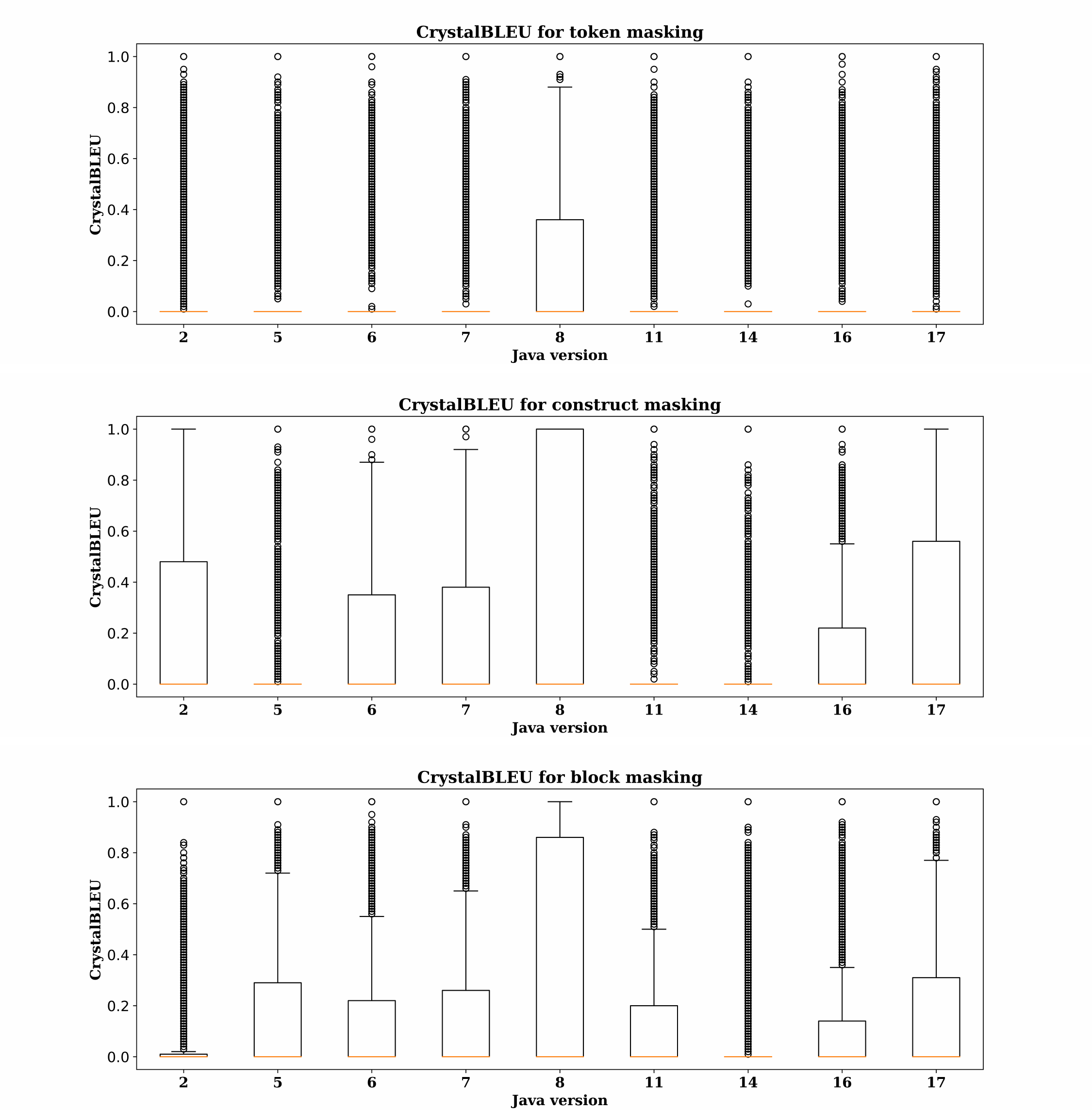}
    \caption{Distribution of the CrystalBLEU score for each Java version and code completion scenario for the non pre-trained model.}
    \label{fig:crystalbleu}
\end{figure*}

\figref{fig:crystalbleu} depicts a boxplot illustrating the distribution of the CrystalBLEU score for each Java version and for each code completion scenario for the non pre-trained model (the same plot for the pre-trained model is available in our replication package \cite{replication}). As illustrated, for the boxes corresponding to Java 8, the 75th percentile and the maximum value are consistently higher than the other Java versions across all completion scenarios. This suggests that the CrystalBLEU score achieves its highest values for Java 8 when compared to the other Java versions, thus indicating superior performance when the model is tested on the same Java version used for training. Since the conclusions using the CrystalBLEU score and EM predictions are aligned, in what follows, we focus the results discussion on the latter metric.

%\begin{figure}[tb]
%    \centering
%    \includegraphics[width=\linewidth]{figures/crystalbleu}
%    \caption{Distribution of the CrystalBLEU score when varying the Java version and the code completion scenario for the non-pretrained model.}
%    \label{fig:crystalbleu}
%\end{figure}

\begin{figure}[tb]
    \centering
    \includegraphics[width=\linewidth]{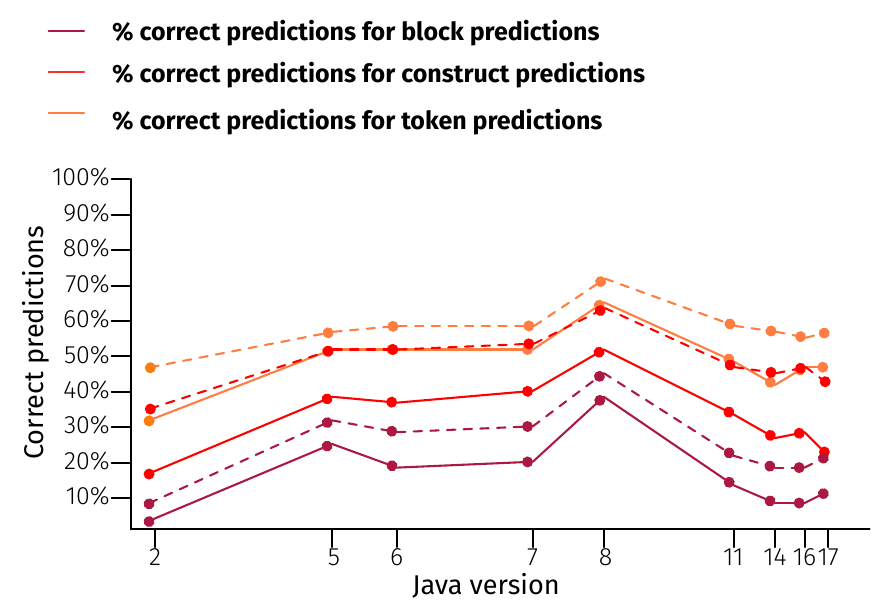}
    \caption{Percentage of Exact Match predictions with and without pre-training on different Java versions.}
    \label{fig:with_vs_without}
\end{figure}

\subsection{Performance Differences}
We computed the percentage of EM predictions for the test set of each Java version, both for the pre-trained and non pre-trained model. \figref{fig:with_vs_without} reports the results obtained.
The $x$-axis shows the Java version, with the distance between data points being proportional to the time passed between Java version releases. The $y$-axis shows the percentage of correct predictions made by the pre-trained (dashed lines) and non pre-trained (continue lines) models in the three masking scenarios, \ie token-level (orange line), construct-level (red), and block-level (dark red).

As illustrated, the trend in performance is similar for different masking scenarios and models. As expected, the best results are obtained for Java 8 (\ie when testing the model on the same Java version seen during training), where the pre-trained model is able to correctly predict up to 70\% of instances for the token-level masking scenario. The percentage of EM predictions gradually decreases as we move away from Java 8, both back and forward in time. Despite the similar trend overall, there are notable differences across language versions and masking scenarios. As expected, block-level masking is the most challenging scenario, since the model has to guess up to several entire statements, while token-level masking is the easiest one, with just a few tokens to predict. Construct-level masking is in the middle: constructs define the application logic and hence are difficult to predict, while still limited in size.

The impact of pre-training is relevant, with an average improvement across all Java versions and masking scenarios of +9\%. 

In details, looking at the performance of the model for each specific completion scenario, the improvement goes from 5\% to 16\% for token-level masking, and from 12\% to 19\% for construct-level masking. For block-level masking the improvement is less evident, ranging from 3\% to 8\% block, likely due to the difficulty of the task.  
Thus, pre-training the model can be very valuable, especially when the predictions are quite challenging, involving for example logic constructs. 

The pre-training dataset is a mixture of all different Java versions~\cite{husain:arxiv2019} and this can help the model in predicting code that, despite being different from the one used for training (fine-tuning), has been seen during the pre-training phase.

In terms of language version, the worst results are obtained for Java 2, probably because this is an archetypal version of the language (Java 2 was released in 1998 while Java 8 in 2014). It is worth noticing that the performance on the last three versions (\ie Java 14, 16 and 17) is very similar, likely due to very small differences between these versions, all released between March 2020 and September 2021.

% !TEX root = ../main.tex

% Design tables. 

   \begin{table}[ht]
\centering
	\caption{Comparing EM predictions on the Java 8 test set \emph{vs} the test sets of the other versions: adjusted $p$-value and OR.}
	 \small
	\label{tab:statistics}
	\begin{tabular}{llrr}
   	\toprule
	{\bf Java} & {\bf Masking} & {\bf Fisher} & {\bf Fisher}\\
	{\bf version} & {\bf scenario} & {\bf p-value} & {\bf OR} \\\midrule
\multirow{3}{*}{2} & token & $\ll0.001$ & 3.74 \\ 
   & construct & $\ll0.001$ & 4.73 \\ 
  & block & $\ll0.001$ & 13.73 \\
 \addlinespace[0.08cm]\hline\addlinespace[0.08cm]
  \multirow{3}{*}{5} & token & $\ll0.001$ & 1.73 \\ 
   & construct & $\ll0.001$ & 1.84 \\ 
   & block & $\ll0.001$ & 1.72 \\ 
  \addlinespace[0.08cm]\hline\addlinespace[0.08cm]
 \multirow{3}{*}{6} & token & $\ll0.001$ & 1.69 \\ 
   & construct & $\ll0.001$ & 1.92 \\ 
   & block & $\ll0.001$ & 2.28 \\ 
  \addlinespace[0.08cm]\hline\addlinespace[0.08cm]
 \multirow{3}{*}{7} & token & $\ll0.001$ & 1.59 \\ 
   & construct & $\ll0.001$ & 1.63 \\ 
   & block & $\ll0.001$ & 2.16 \\ 
  \addlinespace[0.08cm]\hline\addlinespace[0.08cm]
 \multirow{3}{*}{11} & token & $\ll0.001$ & 1.70 \\ 
   & construct & $\ll0.001$ & 1.97 \\ 
   & block & $\ll0.001$ & 2.65 \\ 
  \addlinespace[0.08cm]\hline\addlinespace[0.08cm]
 \multirow{3}{*}{14} & token & $\ll0.001$ & 2.06 \\ 
   & construct & $\ll0.001$ & 2.51 \\ 
   & block & $\ll0.001$ & 3.92 \\ 
 \addlinespace[0.08cm]\hline\addlinespace[0.08cm]
  \multirow{3}{*}{16} & token & $\ll0.001$ & 1.83 \\ 
   & construct & $\ll0.001$ & 2.27 \\ 
   & block & $\ll0.001$ & 3.66 \\ 
 \addlinespace[0.08cm]\hline\addlinespace[0.08cm]
  \multirow{3}{*}{17} & token & $\ll0.001$ & 1.78 \\ 
   & construct & $\ll0.001$ & 3.05 \\ 
   & block & $\ll0.001$ & 2.64 \\ 
\bottomrule
\end{tabular}
\end{table}

We also reported in \tabref{tab:statistics} the results of the Fisher's Exact test (and related OR) when comparing the performance on the Java 8 test set and all other test sets (in terms of EM predictions) for the non pre-trained model (results for the pre-trained one are available in our replication package \cite{replication}).

The $p$-values, after adjustment, are always very close to 0, indicating a statistically significant difference in the performance observed on the different test sets. The OR goes from 1.59 to 13.73 indicating much higher odds of observing a correct prediction on the Java 8 test set rather than on the others. 

For example, in the comparison between Java 8 and Java 14 in the token-level scenario, the OR=2.06 indicates that the odds of an EM prediction for Java 8 are $\sim$2 times higher than for Java 14.

\begin{tcolorbox}[top=5pt,bottom=5pt,left=5pt,right=5pt]
\underline{\textbf{Key insight}}: Regardless of the masking scenario and the use of pre-training, there are significant performance differences across language versions, ranging from 11\% (w.r.t. Java 5, block masking, non pre-trained model) to 37\% (w.r.t. Java 2, block masking, pre-trained model).
\end{tcolorbox}

\subsection{Reasons Behind Performance Differences}

The observed drop in performance provides a strong evidence of the \emph{concept drift}~\cite{gama2014survey,lu2018learning} issue caused by programming language evolution. In this section, we try to better link the observed drop in performance to the changes implemented across the Java versions. In particular, we looked for the official Java documentation reporting the new APIs introduced in each Java version different from Java 8. Unfortunately, we only found this information for the newer versions (\ie those following Java 8), and only for the versions listed in \tabref{tab:new} we identified new APIs. 

\begin{figure*}[tb]
    \centering
    \includegraphics[width=0.62\linewidth]{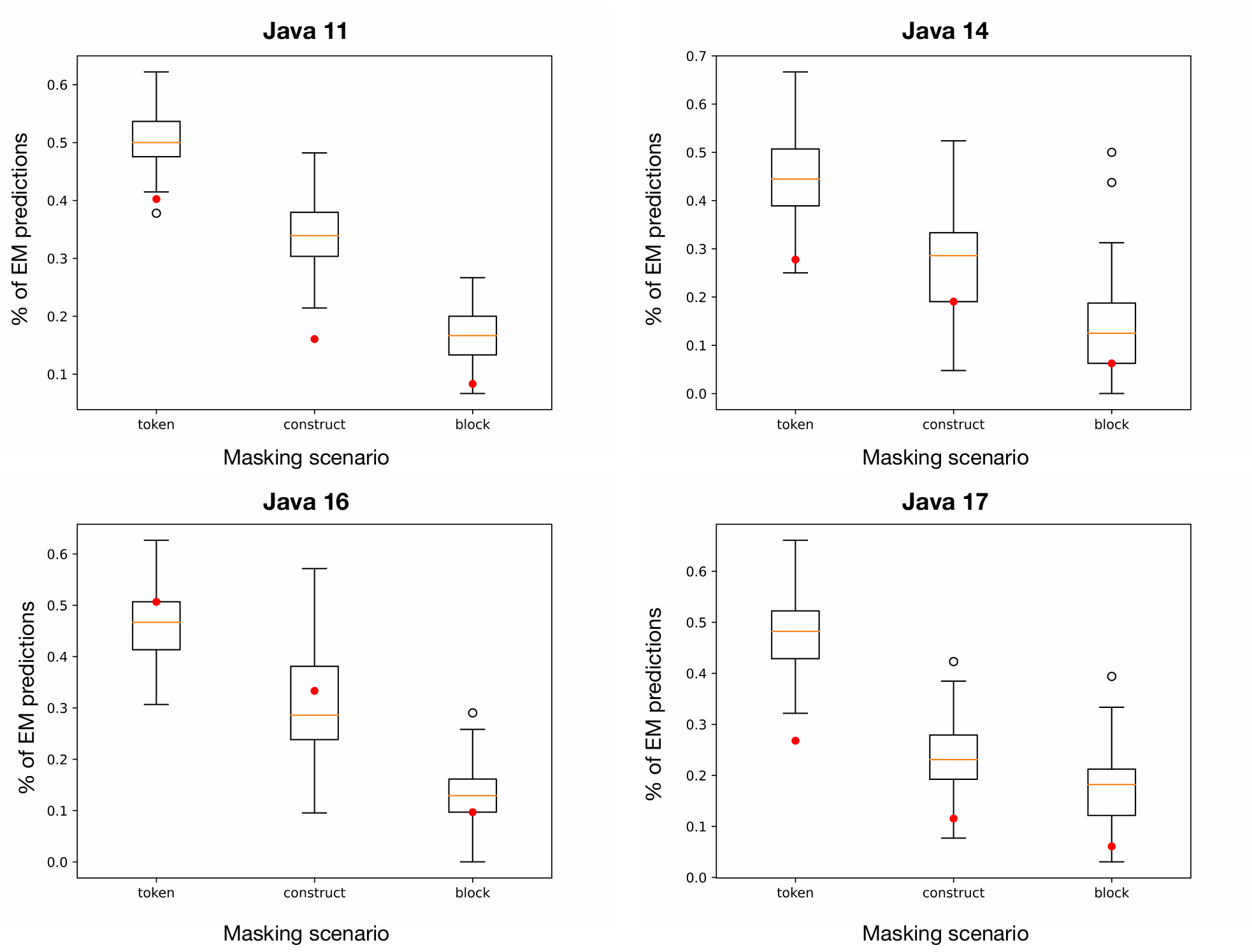}
    \caption{Distribution of the percentage of EM predictions for 100 subsets of instances not featuring new APIs (box plots) \emph{vs} the percentage of EM predictions in the set featuring new APIs (red dots). On average, each set features 43 instances.}
    \vspace{-0.2cm}
    \label{fig:boxplots}
\end{figure*}

% !TEX root = ../main.tex

% Design tables. 

\begin{table}[ht]
	\centering
	\caption{New APIs introduced in specific Java version.}
	 \small
	\label{tab:new}
	\begin{tabular}{ll}
	\toprule
	{\bf Java} & \multirow{2}{*}{\bf Features}\\
	{\bf version} & \\\midrule
	\multirow{2}{*}{9} &  \textit{List.of, Set.of, Map.of, takeWhile, iterate,}\\
	    & \textit{dropWhile, ifPresentOrElse}\\
	    \addlinespace[0.08cm]\hline\addlinespace[0.08cm]
	10 & \textit{List.copyOf, orElseThrow}\\
	\addlinespace[0.08cm]\hline\addlinespace[0.08cm]
	\multirow{3}{*}{11} & \textit{isBlank, lines, strip, stripLeading, stripTailing,}\\
	 & \textit{repeat, Files.writeString, Files.readString, }\\
	 & \textit{HttpClient, HttpRequest, HttpResponse}\\
	 \addlinespace[0.08cm]\hline\addlinespace[0.08cm]
	\multirow{2}{*}{12}  & \textit{indent, transform, Files.mismatch, Collector.teeing, }\\
	 & \textit{NumberFormat.getCompactNumberInstance}\\
	 \addlinespace[0.08cm]\hline\addlinespace[0.08cm]
	\multirow{2}{*}{13}  & \textit{newFileSystem, stripIndent, translateEscapes,} \\
	& \textit{formatted}\\

\bottomrule
\end{tabular} 
\end{table}

Based on this information, we conducted an analysis to compare the percentage of EM predictions in the instances from the test sets containing new APIs against the percentage of EM predictions in the instances not featuring any new APIs released after Java 8. The goal of this analysis is to understand whether the presence of new APIs (not seen in the Java 8 training set) in the code to predict has an impact on the model performance.

It is worth mentioning that the code containing new APIs represents less than 1\% of the instances of the test sets. Thus, the observed drop in performance is certainly not \emph{only} due to the new APIs introduced in the new Java versions. 

For example, Java 9 also introduced an improved \texttt{try-with\--resource} statement and the diamond operator extensions which partially changed the language syntax. Still, we compare the performance of the code completion model when dealing with instances featuring and not featuring new APIs to get an idea of what the impact of new ``code tokens'' unseen in the training set can be. 

To allow for a fair and robust comparison, given $n$ the number of instances featuring new APIs in the test set of Java version $j$ (with $j \in \{11, 14, 16, 17\}$, namely one of the versions following Java 8 and considered in our study) we randomly select from the same test set 100 subsets of $n$ instances each, all not featuring new APIs. Then, we compare the percentage of EM predictions obtained on the set featuring instances with new APIs with the distribution of EM predictions obtained for the 100 subsets of equal size featuring instances not exploiting new APIs.

% Using this information, we conducted an analysis to determine whether the percentage of EM predictions is different in the test set instances featuring the new APIs, namely $D_{new}$ as compared to the test set instances that do not incorporate any new API released after Java 8, namely $D_{no\_new}$. 
% It is worth noting that the new APIs introduced are pretty rare, accounting less than 1\% of the instances, maybe due to the reluctance of developers to experiment with new features, and preventing us from simply comparing the percentage of EM predictions in $D_{new}$ and $D_{no\_new}$.
% To address this challenge, we randomly sampled from $D_{no\_new}$ the same number of instances featuring new APIs and then we compute the percentage of EM predictions on this subset. We iterated this procedure 100 times to bolster the robustness and reliability of our analysis, given the randomness of the sample selection.

Our findings are presented in \figref{fig:boxplots}. For each Java version and masking scenario, we report a boxplot illustrating the distribution of the percentage of EM predictions across the 100 randomly sampled subsets, \ie those not featuring new APIs. We also report a red dot representing the percentage of EM predictions in the instances containing new APIs. As illustrated, for the Java 11, 14, and 17 test sets, the performance observed on the instances featuring the new APIs usually falls below (or inline) with the first quartile, indicating that in $\geq$75\% of cases, we observed better performance in the instances not featuring new APIs. This holds for all code completion scenarios (\ie token, construct, block). The only exception to this trend is Java 16, in particular when dealing with token- and construct-level completions.

\begin{tcolorbox}[top=5pt,bottom=3pt,left=5pt,right=5pt]
\underline{\textbf{Key insight}}:
The introduction of new APIs in the language has, in most cases, an impact on the model performance, with a noticeable drop of EM predictions for instances featuring new APIs. Still, no strong claims can be made on this finding given the lack of statistical analysis, which we do not perform given the low number of instances in the test sets exhibiting new APIs (43, on average, in each of the 12 test sets).

\end{tcolorbox}

\subsection{Impact of Version-Specific Fine-Tuning}
\label{sub:version-specific}
Given the major drop in performance observed when moving away from Java 8, possible strategies to address such a performance decrease are worth being investigated. For this reason, we studied the extent to which a small additional fine-tuning performed on each of the eight Java versions $v_j \neq 8$ may increase performance on the $v_j$ test set. In particular, we created eight additional fine-tuning datasets (one per each Java version different from Java 8) by using methods that have been excluded while building the test sets (\ie excluded by  Algorithm \ref{algorithm}) and thus not used in any step of our study. This resulted in the datasets listed in  \tabref{tab:fine_tuning}. The datasets have a different size, also allowing to observe whether particularly small datasets (\eg Java 17) are anyway sufficient to observe any practical improvement. Each of these datasets has been split into 90\% for the additional fine-tuning and 10\% for evaluation. The fine-tuning has been performed for only five epochs on top of the model fine-tuned for Java 8, in an attempt to simulate its adaptation to a different Java version. In total, 16 new models have been trained (\ie 8 Java versions with/without pre-training). We assessed the EM predictions of the models after each training epoch on the corresponding evaluation set, selecting the best performing one to be run on the test set (\ie the one adopting the same version as the version-specific fine-tuning dataset).

% Design tables. 

\begin{table}[t]
	\centering
	\caption{Number of methods and instances of the version-specific fine-tuning datasets for each Java version.}
	\small
	\label{tab:fine_tuning}
	\begin{tabular}{lrrrr}
	\toprule
	\bf Java & \multirow{2}{*}{\bf \# Methods} & {\bf \# Token} & {\bf \# Construct} & {\bf \# Block}\\ 
	\bf Version & & {\bf Instances} & {\bf Instances}& {\bf Instances} \\\midrule
 2 & 1,530 & 4,475 & 510 & 750\\
 5 & 3,549 & 8,552 & 1,840 & 4,237\\
 6 & 63,379 & 168,605 & 49,959 & 79,123\\
 7 & 94,405 & 252,722 & 68,561 & 22,124\\
 11 & 92,153 & 105,700 & 33,111 & 38,982\\
 14 & 1,160 & 3,348 & 1,664 & 1,894\\
 16 & 2,169 & 6,237 & 1,378 & 2,956\\
 17 & 915 & 2,493 & 576 & 1,073\\\midrule
 \bf ALL & \bf 259,260 & \bf 552,132 & \bf 157,599 & \bf 151,139\\
\bottomrule
\end{tabular}
\end{table}

\begin{figure*}[t]
    \centering
    \includegraphics[width=0.9\linewidth]{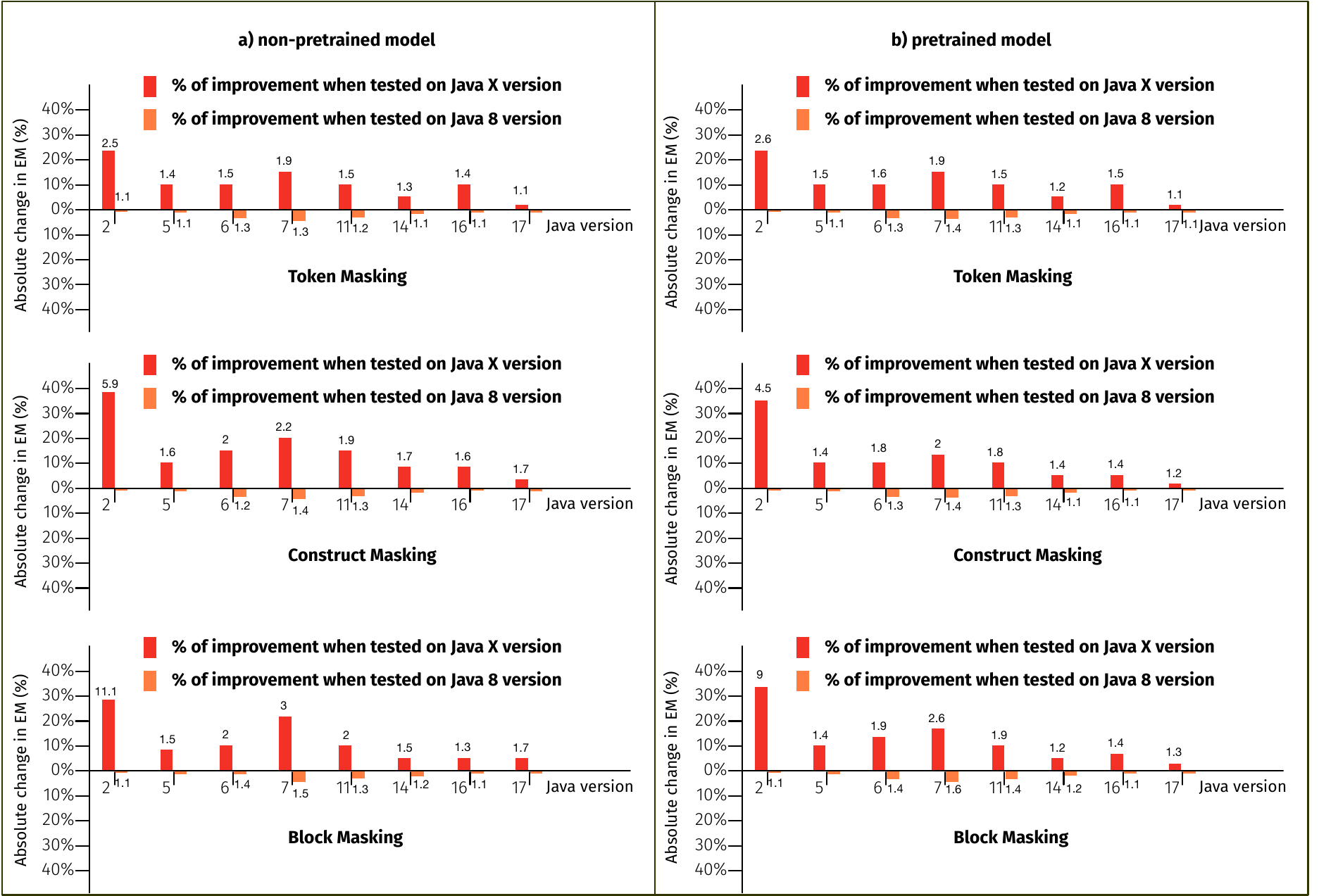}
    \caption{Difference in the percentage of EM predictions with and without the version-specific fine-tuning, evaluated on the same Java version test set and on the Java 8 test set.}
    \label{fig:specific}
\end{figure*}

\figref{fig:specific} reports the achieved results for the non pre-trained (a) and for the pre-trained (b) models. Each subfigure features eight pairs of bars, one pair for each of the eight Java versions for which we further fine-tuned the model (\ie all but Java 8). The red bars represent the absolute improvement in EM predictions observed on the test set adopting the same version used for further fine-tuning the model, while the orange bars report changes in performance observed on the Java 8 test set. 
% Indeed, it is possible that in an attempt to adapt the model to different Java versions, the performance on the ``original'' version of interest (\ie Java 8) may degrade.

We computed the performance on both test sets in an attempt to understand whether the adaptation of the model to a specific Java version (different from Java 8) may have a negative impact on the version that the model was originally trained on (Java 8).

% \MATTEO{To assess the significativity of the improvement, we computed the Fisher's exact test (and related OR) comparing the performance with and without the version-specific fine-tuning, on both the test sets described before. We reported on the top of each bar the OR when the test is statistically significant. }
We statistically analyzed the results using the Fisher's exact test and related OR, comparing the performance of the model with and without the version-specific fine-tuning for each of the test sets previously described.
% We reported on the top of each bar the OR when the test is statistically significant.
For the statistically significant comparisons (\ie $p$-value $<$ 0.05), we reported the OR value on the top of each bar. Since the trend is similar for both pre-trained and non pre-trained model, we focus our discussion on the non pre-trained one).

% \GABRIELE{Revise once new figure is ready}
The second round of fine-tuning significantly increased the accuracy of the predictions. The most notable improvement was obtained for Java 2---the most negatively affected version by the concept drift problem---where
% (where the lowest performance was achieved before the second fine-tuning):
the percentage of correct block-level predictions jumped from 3\% to 33\%, with an overall improvement of 31\% across all masking scenarios. All improvements resulted in statistically significant differences with large ORs (\ie 2.5 in the token-level, 5.9 in the construct-level, and 11.1 in the block-level). 

For the other Java versions, the improvement ranges from 3\% to 15\% for token-level predictions, from 7\% to 39\% for construct-level predictions, and from 4\% to 29\% for block-level predictions. Overall, the average improvement is 11\% across all masking scenarios. Interestingly, even a rather small fine-tuning dataset such as the one used for Java 17 still allowed to achieve statistically significant better performance in all masking scenarios, although with ORs limited to a maximum of 1.7.

Such a major improvement has however a small price to pay. Indeed, re-adapting the model to a specific version did have an impact on the performance of the model for the version it was originally trained on (Java 8). However, as shown in Figure~\ref{fig:specific} (orange bars), the performance drop was very small in most cases (4\% on average).
These findings show that a short additional fine-tuning on a specific Java version can significantly improve the performance of the model on that version, while incurring a negligible performance degradation for the original version the model was trained on.

\begin{tcolorbox}[top=5pt,bottom=5pt,left=5pt,right=5pt]
\underline{\textbf{Key insight}}: A limited fine-tuning---with few training instances and epochs---on a specific language version can lead to significant performance improvements in the model predictions (up to 40\%) at the cost of a negligible performance drop in the original version (4\% on average). The most noticeable improvements are achieved for the most challenging tasks, \ie construct-level and block-level predictions.
\end{tcolorbox}

% !TEX root = main.tex
%%%%%%%%%%%%%%%%%%%%%%%%%%%%%%%%%%%%%%%%
%%%%%%%%%%%%%%%%%%%%%%%%%%%%%%%%%%%%%%%%
\section{Threats to Validity} \label{sec:threats}
%%%%%%%%%%%%%%%%%%%%%%%%%%%%%%%%%%%%%%%%
%%%%%%%%%%%%%%%%%%%%%%%%%%%%%%%%%%%%%%%%

% \alberto{
% Possible things to discuss:
% \begin{itemize}
%     \item Different sizes of test sets.
%     \item External validity: would the conclusions hold for different DL models, programming languages, or if selecting a different Java version for the training set?
%     \item For the blocks dataset, only blocks containing at most three statements are masked. Would we observe similar performance differences if we had selected longer blocks?
%     \item For the ``impact of version-specific fine-tuning'', the dataset sizes reported in \tabref{tab:fine_tuning} are very different. The differences in performance observed across versions may be due to this.
%     \item Measuring performance in terms of ``Exact Match'' can be misleading: sometimes, even if the match is not exact, the prediction is still correct (\eg you can use different Java APIs to achieve the same result).
%     \item The test sets may be different, which may explain performance differences, but we used Algorithm 1 to create fairly similar sets.
% \end{itemize}
% }

\textbf{Conclusion validity}. We applied  appropriate statistical analysis, using specific statistical tests and effect size measures, following common guidelines in the literature~\cite{arcuri2014hitchhiker,yoav:jstor1995}.

\textbf{Construct validity}. These revolve mainly around the way in which we evaluate the task of code completion, \ie by masking code elements. While we acknowledge that this might not be completely representative of how developers write code, we evaluate three different completion tasks (token-level, construct-level and block-level) that allow us to evaluate the model performance in different scenarios and with different amounts of masked code. Another threat in this regard is \emph{how} we evaluate the model performance, \ie by measuring the percentage of \emph{Exact Matches} and the CrystalBLEU score. While it is possible that the model predicts different but semantically equivalent code, we believe that these metrics are still a good proxy for the model performance, as previous work shows that distinct code suggestions tend to be semantically different~\cite{Ciniselli:msr2021,asaduzzaman:icsme2014,alon:icml2020,hellendoorn:fse2017}. %\alberto{Do you know of any other reference to support this last statement? I think it's important to put more than one example.}

\textbf{Internal validity}. On the one hand, it is possible that the observed performance differences across language versions may be due to the specific test sets used for the experiments. To partially address this threat, we implemented Algorithm~\ref{algorithm}, whose goal is to balance the complexity of the predictions in the tests. On the other hand, the impact of version-specific fine-tuning (Section~\ref{sub:version-specific}) might depend on the size of the fine-tuning datasets. In acknowledging this, we reported the sizes of all version-specific datasets used (Table~\ref{tab:fine_tuning}). %As a matter of fact, this also helps understand the impact of the dataset size on the fine-tuning itself (Figure~\ref{fig:specific}).

\textbf{External validity}. %Threats to the external validity may affect the generalizability of our findings.
% The main threat in this regard is the specific features characterizing our study, namely: (i) CodeT5 as the selected DL model; (ii) Java as the selected programming language; and (iii) Java 8 as the selected reference language version against which to compare the other versions.
% The features that characterize our study are: (i) the DL model, CodeT5; (ii) the programming language, Java; and (iii) the reference language version, Java 8.
% While it is possible that the results of our study may not generalize to other settings, we believe that CodeT5 and Java are a good representative of popular, widely used DL models and languages, respectively. Moreover, the choice of Java 8 as the reference language is motivated by the 
Our study is characterized by the selected DL model, programming language, and reference language version. CodeT5 was selected as a representative code model that has demonstrated strong performance across a range of tasks~\cite{wang:emnlp2021}. The choice of Java was motivated by its popularity and widespread use, and because it is possible to reliably identify a project's Java version based on its \texttt{POM} file. Finally, we selected Java 8 as the reference version due to its prevalence in the dataset, as shown in Table~\ref{tab:dataset}.
% All in all, the magnitude of the experiments (over one million methods from 784 Java repositories) make us remain confident of the significance of the results.

% !TEX root = main.tex
 
%%%%%%%%%%%%%%%%%%%%%%%%%%%%%%%%%%%%%%%%
%%%%%%%%%%%%%%%%%%%%%%%%%%%%%%%%%%%%%%%%
\section{Related Work} \label{sec:related}
%%%%%%%%%%%%%%%%%%%%%%%%%%%%%%%%%%%%%%%%
%%%%%%%%%%%%%%%%%%%%%%%%%%%%%%%%%%%%%%%%
% \alberto{@Matteo, in both subsections, I am missing some kind of statement (one or more, interleaved with the text or at the end) where we compare our work against the others, highlighting the differences.}
% \MATTEO{I though about that but this work is extremely different from the reported works (the first section aims to describe different approaches while in the second section the authors mostly investigate the way in which developers use code recommenders). I would avoid that but feel free to add something if you have any idea.}

Our work is related to ML and DL models for code completion, as well as empirical studies investigating the usage of DL-based solutions for code completion. %In what follows, we discuss the most relevant work in these two areas.

\subsection{Machine and Deep Learning Models for Code Completion}

One of the first techniques proposed in the literature to support code completion is the Prospector of Mandelin \etal \cite{mandelin:pldi2005}. The tool is able to recommend IDE variables and method calls.

Bruch \etal~\cite{bruch:fse2009} proposed the Best-Matching Neighbor (BMN) techniques to recommend a method call. Their approach leverages method usage frequency and association rules to suggest the calls that  are more relevant to the code written in the IDE, achieving 82\% of precision and 72\% of recall.

Robbes and Lanza~\cite{robbes:ase2010} exploited the history of software systems to recommend method calls and class names. Their approach has been implemented in a tool named OCompletion, showing a top-3 accuracy of 75\%.

Asaduzzaman \etal~\cite{asaduzzaman:icsme2014} introduced CSCC (Context Sensitive Code Completion). CSCC creates a context for each method call composed of methods, keywords, classes, and interfaces appearing within four lines from the call. Thanks to the improved context, they achieved 86\% precision and 99\% recall in call completion.

Hindle \etal~\cite{hindle:icse2012} applied the $n$-gram statistical language model to the prediction of the next token in a given statement. They hypothesized the ``naturalness of source code'' conjecturing that, since the code has been written by humans, it tends to be repetitive and predictable, similarly to natural language. Tu \etal~\cite{tu:fse2014} and Hellendoorn and Devanbu~\cite{hellendoorn:fse2017} proposed  improvements to the $n$-gram model, adding new components to exploit local information, since the code tends to be locally repetitive and specific.

With the rise of DL models, different code completion approaches have been proposed.
Kim \etal~\cite{kim:icse2021} used the Transformer model for code completion, exploiting the Abstract Syntax Tree to strengthen the self-attention with syntactic information. Similarly, Alon \etal~\cite{alon:icml2020} leveraged the syntax of the code to propose the Structural Language Model which combines LSTMs and Transformers.

Differently from the previous two approaches, Svyatkovskiy \etal~\cite{svyatkovskiy:fse2020} did not incorporate any syntactical information in their Transformer-based model, named IntelliCode Compose. Their model, trained on a multi-lingual dataset, was able to predict entire sequences of tokens, showing an unprecedented perplexity of 1.82 when predicting Python tokens.

Ciniselli \etal~\cite{ciniselli:tse2021} performed an empirical analysis assessing the performance of T5 and RoBERTa models when predicting up to two entire statements, comparing their results with the $n$-gram model proposed by Hellendoorn and Devanbu~\cite{hellendoorn:fse2017}.

They showed the superiority of the T5 model, with an accuracy ranging from 29\% up to 69\%, depending on the complexity of the prediction (longer predictions are more challenging).

Feng \etal \cite{feng:emnlp2020} proposed CodeBERT, a Transformer trained on code and English text, able to capture semantic connections between natural and programming language, achieving state-of-the-art performance in different code-related tasks. The authors introduced a novel training objective, the replaced token detection, in which the model has to detect the token that has been replaced with a plausible alternative. 

Wang \etal \cite{wang:emnlp2021} presented CodeT5, an improved version of the T5 model~\cite{raffel:jmlr2019} adapted for code, thanks to the new identifiers-aware task able to exploit semantic code information. Their model achieved state-of-the-art performance on the CodeXGLUE benchmark~\cite{lu:nips2021}. Wang \etal \cite{wang:arxiv2023} subsequently extended their previous work presenting CodeT5+, a family of encoder-decoder models where the modules can be easily combined to handle specific tasks, thanks to a plethora of pre-training objectives.

Chen \etal \cite{chen:arxiv2021}  introduced Copilot, a new Transformer model trained on more than 150GB of data from GitHub. They evaluated the model performance by checking whether the proposed solution passes a suite of test cases, showing that standard match-based metrics like the BLEU score are not well suited for measuring the model accuracy. Their trained model achieved state-of-the-art performance in the demanding task of predicting the entire body of a method starting from the natural language description of the task to implement.

All the works previously described focus on the problem of code completion, presenting approaches and techniques to improve this task. Differently, our work evaluates the performance of a state-of-the-art code model, CodeT5, on the task of code completion in a different setup, namely, predicting code belonging to different language versions. In this sense, our study may be complementary to the ones presenting new approaches, since it can guide future solutions tailored at limiting the drop in performance we observed as the programming language evolves.

\subsection{Empirical Studies Investigating the Usage of DL-Based Solutions for Code Completion}

Several studies investigated the usage of DL-based solutions for code completion, highlighting different limitations that can prevent developers from using code recommenders, also suggesting possible ways to improve them.

Hellendoorn \etal \cite{hellendoorn:icse2019} showed that the benchmarks used for code completion tasks are not representative of real code completion tasks. Indeed, the tools they experimented with are less accurate on real-world datasets and often fail in challenging scenarios, when predicting a high number of tokens.

Ciniselli \etal \cite{ciniselli:msr2022} investigated whether DL-based code recommenders tend to suggest code that is copied from the training set. Their findings showed that $\sim$10\% of the recommendations are code clones, although the copied snippets are often short and very popular suggestions, like {\tt return} statements.   

M{\u{a}}r{\u{a}}șoiu \etal \cite{muaruasoiu:ppig2015} investigated the usage of code completion tools in practice, observing that most of the times developers discard the recommendations.

The main reason for the low acceptance rate lies in the limited familiarity of the developer with some APIs, pushing the developer to an unsuccessful exploration of the recommender's suggestions.

Jin and Servant \cite{jin:msr2018} examined the \textit{hidden costs} of code recommendations. They found that the productivity of the developer decreases when the tool provides a high number of suggestions, discouraging them from using code recommenders.

Liu \etal \cite{liu:tse2020} assessed the capabilities of DL-based models in generating code starting from requirement texts. They curated a new dataset using real-world tasks found in online programming contest platforms, showing significantly lower performance as compared to the standard benchmarks reported in the literature. The authors hypothesized that the main reason behind this drop is the small size of the datasets used for the model evaluation.

Xu \etal \cite{xu:tosem2022} performed an empirical study in which developers had to implement novel tasks with and without the support of code recommenders. Their findings showed that these tools are not able to increase the developers' productivity.

Ziegler \etal \cite{ziegler:maps2022} attempted to define a quantitative metric that better aligns with the perceived productivity of developers when using GitHub Copilot. Results showed that the acceptance rate of the suggestions proposed by the tool is highly correlated with the developers' productivity.

Compared to these studies, our work investigates a new problem not previously addressed: the impact of the concept drift caused by language evolution on the performance of DL-based code completion approaches.

% !TEX root = main.tex
%%%%%%%%%%%%%%%%%%%%%%%%%%%%%%%%%%%%%%%%
%%%%%%%%%%%%%%%%%%%%%%%%%%%%%%%%%%%%%%%%
\section{Conclusion and Future Work} \label{sec:conclusion}
%%%%%%%%%%%%%%%%%%%%%%%%%%%%%%%%%%%%%%%%
%%%%%%%%%%%%%%%%%%%%%%%%%%%%%%%%%%%%%%%%
Programming languages evolve rapidly. To support developers effectively, coding tools must keep up with this fast pace. DL-based code completion approaches and, more broadly, intelligent coding assistants such as the recently released GitHub Copilot Chat~\cite{githubcopilotchat} are inevitably affected by these changes. In this paper, we investigated the impact of language evolution on the performance of code models for the task of code completion.

Our research provides \emph{actionable results} for researchers, developers and tool builders. With respect to \emph{researchers}, we provide solid empirical evidence about the notion of concept drift when it comes to the evolution of programming languages. This also implies that results achieved on a specific language version cannot be generalized to other (different) versions. Concerning \emph{developers}, they can expect a drop of quality in the recommendations of the code completion engine they use when new versions of a language are released. Finally, \emph{tool builders} cannot just deploy trained models without maintaining them over time, but they need to re-train the models as soon as new versions of the supported programming language(s) become popular. In this sense, our preliminary evaluation of the impact of version-specific fine-tuning is encouraging, as it shows that even a small amount of version-specific data (\eg a few hundred samples) can significantly improve the performance of the model.

Our future research will target two main goals. First, we plan to make our study more comprehensive and extend it to other DL models and programming languages. Second, we will delve deeper into the impact of the small version-specific fine-tuning on the model performance, as well as the perceived improvements by developers in real-world scenarios. These efforts will contribute to the development of adaptable code completion tools, ensuring they remain effective amid the dynamic nature of programming languages.

\section*{Acknowledgment}
This project has received funding from the European Research Council (ERC) under the European Union's Horizon 2020 research and innovation programme (grant agreement No. 851720).

\balance

\bibliographystyle{ACM-Reference-Format}
\bibliography{main}
\newpage

\end{document}